\documentclass[aps,prl,twocolumn,superscriptaddress]{revtex4-1}

\usepackage{graphicx}

\begin{document}

\title{Triplet and in-gap magnetic states in the ground state of the quantum frustrated FCC antiferromagnet Ba$_2$YMoO$_6$}

\author{J.~P.~Carlo}
\affiliation{Department of Physics and Astronomy, McMaster University, Hamilton, ON  L8S 4M1  Canada}
\affiliation{Canadian Neutron Beam Centre, National Research Council, Chalk River, ON K0J 1J0  Canada}

\author{J.~P.~Clancy}
\affiliation{Department of Physics and Astronomy, McMaster University, Hamilton, ON  L8S 4M1  Canada}

\author{T.~Aharen}
\affiliation{Department of Chemistry, McMaster University, Hamilton, ON L8S 4M1 Canada}

\author{Z.~Yamani}
\affiliation{Canadian Neutron Beam Centre, National Research Council, Chalk River, ON K0J 1J0  Canada}

\author{J.~P.~C.~Ruff}
\affiliation{Department of Physics and Astronomy, McMaster University, Hamilton, ON  L8S 4M1  Canada}

\author{J.~Wagman}
\affiliation{Department of Physics and Astronomy, McMaster University, Hamilton, ON  L8S 4M1  Canada}

\author{G.~J.~{Van~Gastel}}
\affiliation{Department of Physics and Astronomy, McMaster University, Hamilton, ON  L8S 4M1  Canada}

\author{H.~M.~L.~Noad}
\affiliation{Department of Physics and Astronomy, McMaster University, Hamilton, ON  L8S 4M1  Canada}

\author{G.~E.~Granroth}
\affiliation{Spallation Neutron Source, Oak Ridge National Laboratory, Oak Ridge, TN 37831  USA}

\author{J.~E.~Greedan}
\affiliation{Department of Chemistry, McMaster University, Hamilton, ON L8S 4M1 Canada}
\affiliation{Brockhouse Institute for Materials Research, McMaster University, Hamilton, ON L8S 4M1 Canada}

\author{H.~A.~Dabkowska}
\affiliation{Brockhouse Institute for Materials Research, McMaster University, Hamilton, ON L8S 4M1 Canada}

\author{B.~D.~Gaulin}
\affiliation{Department of Physics and Astronomy, McMaster University, Hamilton, ON  L8S 4M1  Canada}
\affiliation{Brockhouse Institute for Materials Research, McMaster University, Hamilton, ON L8S 4M1 Canada}
\affiliation{Canadian Institute for Advanced Research, Toronto, ON M5G 1Z8 Canada}

\date{\today}

\begin{abstract}
The geometrically frustrated double perovskite Ba$_2$YMoO$_6$ is characterized by quantum $s$=1/2 spins at the Mo$^{5+}$ sites of an undistorted face-centered cubic (FCC) lattice.  Previous low-temperature characterization revealed an absence of static long-range magnetic order and suggested a non-magnetic spin-singlet ground state.    We report new time-of-flight and triple-axis neutron spectroscopy of Ba$_2$YMoO$_6$  that shows a 28 meV spin excitation with a bandwidth of $\sim$4 meV, which vanishes above $\sim$125 K.  We identify this as the singlet-triplet excitation out of a singlet ground state, and further identify a weaker continuum of magnetic states within the gap, reminiscent of spin-polaron states arising due to weak disorder. 
\end{abstract}

\pacs{}

\maketitle


Geometrically frustrated magnetic materials \cite{lacroix, greedan_frustration} are of great topical interest due to the complex interplay between competing interactions resulting in rich phase diagrams, including spin-glass, spin-ice and spin-liquid ground states.  Triangular and tetrahedral architectures are most often associated with geometric frustration, although the phenomenon occurs in diverse systems with various lattices, magnetic interactions, and anisotropies.  In two dimensions (2D), networks of edge- and corner-sharing triangles give rise to the triangular and kagome lattices, respectively, while in three dimensions (3D), tetrahedral networks form the face-centered cubic (FCC) and pyrochlore lattices.

Frustrated lattices of antiferromagnetically (AF) coupled moments have been studied in a variety of materials.  Well-studied 2D systems, consisting of loosely-coupled stacks of planes, include the triangular magnets NaCrO$_2$ \cite{NaCrO2} and VCl$_2$ \cite{VCl2}, Kagome magnets such as Herbertsmithite \cite{herbertsmithite} and several Jarosite AFs (\textit{e.g.} \cite{jarosite}).  Other quasi-2D magnetic materials and models which possess competing interactions exist, with resulting physics very similar to that originating from geometrical frustration, including the so-called $J_1$$-$$J_2$ systems \cite{kageyama_square_lattice}, square planar lattices decorated by magnetic moments with opposing nearest-neighbor and next-nearest-neighbor interactions.  One such system of topical interest is SrCu$_2$(BO$_3$)$_2$ \cite{kageyama_SCBO, gaulin_SCBO}, an experimental realization of the Shastry-Sutherland  $s$=$^1$$\!$/$_2$ Heisenberg model \cite{shastry_sutherland}, with moments on a planar lattice of orthogonally oriented dimers.  Each dimer, composed of two $s$=$^1$$\!$/$_2$ Cu$^{2+}$ moments, exhibits a singlet ground state with an $s$=1 triplet excitation above a $\Delta \sim$ 3~meV gap.  In 3D, well-studied frustrated systems include the rare earth titanates, in which magnetic moments reside on essentially perfect pyrochlore lattices \cite{gardner_pyrochlores_review}, and exhibit a wide variety of ground states including spin ice \cite{harris_HoTiO, gaulin_HoTiO}, long-range order (LRO) \cite{champion_ErTiO, gaulin_ErTiO}, field-induced order \cite{gaulin_YbTiO} and spin liquid \cite{gardner_TbTiO, mirebeau_TbTiO}.   Both classical and quantum spins decorating these lattices have been, and are, of interest.  But the quantum versions can give rise to exotic disordered spin-liquid states, as may be relevant to resonating valence bond states \cite{anderson_rvb}. 

While experimental and theoretical works on classical and quantum quasi-2D triangular and kagome magnets and 3D pyrochlore magnets abound, there are very few studies of quantum FCC frustrated systems.   In rock-salt ordered double perovskites \cite{anderson_dblperovskites} (Fig.~1(a)) the magnetic moments comprise an edge-sharing tetrahedral network (Fig.~1(c)).  While most are not perfect $s$=$^1$$\!$/$_2$ FCC systems (\textit{e.g.}~\cite{diep_fcc, BHSO}), experimental studies have revealed a wealth of ground states.   The $s$=$^3$$\!$/$_2$ systems La$_2$LiRuO$_6$ and Ba$_2$YRuO$_6$ exhibit AF LRO \cite{tomoko_3_2}.  Analogous $s$=1 systems show spin freezing without LRO in Ba$_2$YReO$_6$, and a collective singlet state in La$_2$LiReO$_6$ \cite{tomoko_1}.  The extreme quantum $s$=$^1$$\!$/$_2$ case is realized in Sr$_2$CaReO$_6$ \cite{wiebe_SCRO}, La$_2$LiMoO$_6$ and Ba$_2$YMoO$_6$ \cite{tomoko_1_2}.   While the first two exhibit short-range magnetic correlations without LRO, only Ba$_2$YMoO$_6$ maintains cubic symmetry to 2~K and shows no signs of magnetic order in NMR, muon spin relaxation, neutron diffraction or susceptibility measurements \cite{tomoko_1_2, cussen_BYMO, deVries_VBglass_BYMO}, making it an excellent realization of a quantum FCC antiferromagnet.  

 \begin{figure}[ht]
\includegraphics[width=90mm]{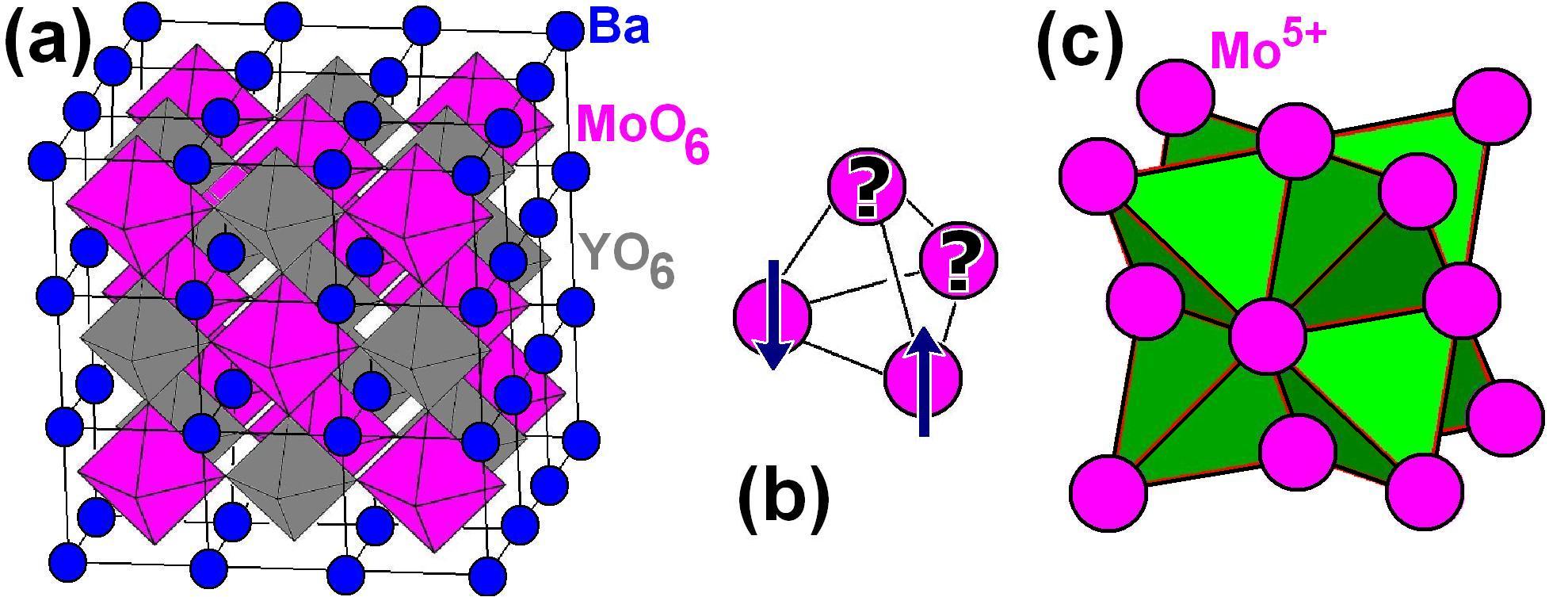}
 \caption{\label{}(a) Unit cell of Ba$_2$YMoO$_6$, with Y and Mo occupying alternate B sites. (b) Schematic diagram depicting frustration of AF-coupled moments on a tetrahedron, and possible formation of orthogonal singlet dimers. (c) Mo$^{5+}$ ions in the lattice in (a), forming a network of edge-sharing tetrahedra.}
 \end{figure}

Ba$_2$YMoO$_6$ was characterized in depth by Aharen \textit{et~al.}~\cite{tomoko_1_2}  Y and Mo ions lie on alternate B sites in an NaCl-like arrangement with only $\sim$3\% B-site disorder, so that the magnetic Mo$^{5+}$ ions form a lattice of edge-sharing tetrahedra.  Bulk susceptibility measurements show high-temperature AF C-W behavior with $\Theta_W$~=~$-$219~K, and some deviation from C-W at lower temperatures.  However, susceptibility, heat capacity and muon spin relaxation measurements found no evidence for a magnetic phase transition above 2~K.  $^{89}$Y NMR $1/T_1$ measurements find two characteristic environments, one corresponding to a paramagnetic-like state at all temperatures, and another indicative of a collective singlet ground state with an effective gap $\Delta$/$k_B$ $\sim$ 140~K.

These results indicate that Ba$_2$YMoO$_6$ exhibits both geometric frustration and strong quantum effects, with a singlet or singlet-like ground state caused by pairing of adjacent $s$=$^1$$\!$/$_2$ Mo$^{5+}$ moments coexisting with a disordered state to 2~K.  Theoretically, Chen \textit{et~al.}~\cite{balents_new} have considered the ground states of $s$=$^1$$\!$/$_2$ FCC systems with strong spin-orbit coupling, as may be expected for 4$d^1$ Mo$^{5+}$ ions.  The spins can combine with the three-fold degeneracy of the $t_{2g}$ orbitals to give an effective  $j$=$^3$$\!$/$_2$  system, allowing a rich variety of exotic ground states, such as a quadrupolar ordered state with spontaneous anisotropy.  But in the case of strong AF interactions pseudo-singlets can arise and lead to non-magnetic valence bond solid and quantum spin liquid ground states.   


In this Letter, we report inelastic neutron scattering results on polycrystalline Ba$_2$YMoO$_6$.  We find scattering at low $Q$ and $E$ $\sim$ 28 meV whose intensity decreases with increasing  temperature and disappears above $\sim$125~K, as well as a continuum of low-$Q$ scattering within this 28~meV gap.  This continuum is weakly peaked in energy and resembles so-called spin polarons, scattering from impurity-induced paramagnetic regions embedded in a sea of singlets \cite{Haravifard_SCBO}.  The $\sim$4 meV bandwidth of the 28~meV spin excitation is consistent with weakly dispersive triplet excitations from a singlet ground state formed from orthogonal dimers on the $s$=$^1$$\!$/$_2$ Mo$^{5+}$ tetrahedra.

Two 6-7g powder samples of Ba$_2$YMoO$_6$ were prepared by conventional solid-state reaction as in \cite{tomoko_1_2}.  A stoichiometric mixture of BaCO$_3$, Y$_2$O$_3$ and MoO$_3$ was fired at 950$^{\circ}$C for 12h, then re-ground and fired at 1250-1300$^{\circ}$C in a reducing 5\% H$_2$/Ar mixture.  Phase purity and the Mo oxidation state were verified through x-ray diffraction and thermogravimetric analysis, respectively.

Measurements were performed on one sample at the SEQUOIA Fine Resolution Fermi Chopper Spectrometer at the Spallation Neutron Source (SNS), Oak Ridge National Laboratory \cite{granroth1, granroth2}, and on the other at the C5 triple-axis spectrometer at the Canadian Neutron Beam Centre (CNBC), Chalk River.   Each specimen was contained in an Al sample can in a closed-cycle refrigerator with He exchange gas, with measurements made on identical empty sample cans for background subtraction.

Time-of-flight measurements at SEQUOIA were performed between 6~K and 290~K, employing an incident beam energy $E_i$ =  60~meV chosen by Fermi chopper \#1 \cite{granroth2} spinning at 240~Hz ($\Delta$$E$/$E\sim$5\%).  Background from the prompt pulse was removed by the $T_0$ chopper at 60~Hz.  The beam was masked to match the sample size, and normalization to a white-beam vanadium run corrected for the detector efficiencies.

Triple-axis measurements at C5 employed pyrolitic graphite (PG) as both monochromator and analyzer, in constant $E_f$ mode using $E_f$ = 30.5 meV, at temperatures from 3.1 to 300~K.  Harmonic contamination in the scattered beam was suppressed using a PG filter.   Collimations along the beam path were [33$'$-47$'$-51$'$-144$'$], with an energy resolution of 4~meV at the elastic channel.  


\begin{figure}[ht]
 \includegraphics[width=85mm]{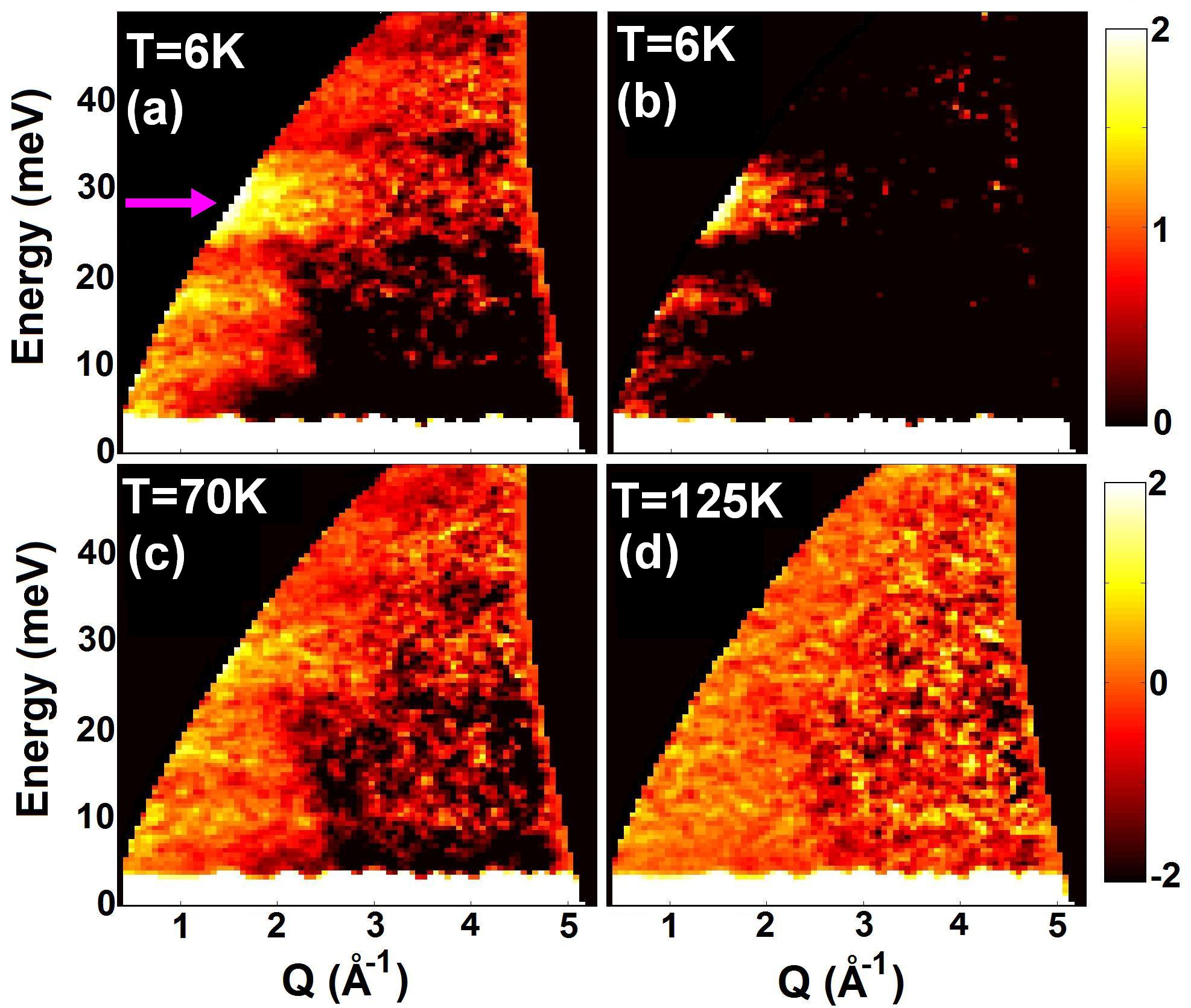}
 \caption{\label{} (a) (c) (d) Dynamic susceptibility,  $\Delta\chi''$(Q,$\hbar\omega$), at $T$~=~6~K, ~70~K, \&~125~K, where  $\chi''$($Q, \hbar\omega$) at $T$~=~175~K has been subtracted from each to isolate the magnetic scattering, as described in the text.  (b) shows  $\Delta\chi''$(Q,$\hbar\omega$), at $T$~=~6~K with $T$=~175~K subtracted, but with the plotted intensity scale range restricted to $>$0 only, thus highlighting where  $\chi''$(Q,$\hbar\omega$), at 6~K exceeds that at 175~K.  The lower intensity scale refers to (a)(c)(d), and the upper refers to (b).}
 \end{figure}

The neutron scattering cross section due to phonons scales as (\textbf{$\vec{\epsilon}\cdot \vec{Q}$})$^2$, where \textbf{$\vec{\epsilon}$} is the phonon eigenvector \cite{squires}, while that from magnetism scales with the form factor of the appropriate magnetic electrons, and generally falls off with increasing $Q$.  To isolate the magnetic scattering, we follow a similar approach to that of Clancy \textit{et~al.} \cite{clancy_TiOBr}, wherein the total scattering intensity is treated as a sum of three factors: a temperature-independent background, a phonon contribution whose temperature dependence is described well by the thermal occupancy factor $[n(\omega)+1]$, and the magnetic contribution of present interest.  The temperature-independent background is removed by subtracting the empty sample-can run from each data set.  This gives $S$($Q, \hbar\omega$), which is normalized by the thermal occupancy factor to yield $\chi''$($Q,\hbar\omega$).   Finally, we subtract $\chi''$($Q, \hbar\omega$) at 175 K from that at 6 K and the other data sets of interest to remove the phonon contribution and approximately isolate the magnetic contribution, resulting in the $\Delta$$\chi''$($Q, \hbar\omega$) maps shown in Fig. 2. Figs. 2(a), (c) and (d) show $\Delta$$\chi''$($Q, \hbar\omega$) maps for $T$=6~K, 70~K and 125~K, with $T$=175~K subtracted on a full intensity scale, while Fig. 2(b) shows the $T$=6$-$175~K subtraction for positive values only, to highlight where $\chi''$($Q, \hbar\omega$) at $T$=6~K exceeds that at $T$=175~K.   Magnetic scattering identified in this way is clearly seen at low Q$<$2.5 $\text{\AA}^{-1}$.  Figs. 2(a), (c) and (d) show $\Delta$$\chi''$($Q, \hbar\omega$) to evolve from a flat $Q$-$\hbar\omega$ distribution to one characterized by two inelastic distributions at small $Q$.  The strongest of these is centered on an energy of 28~meV, with a bandwidth $\sim$4~meV.  The other is a continuum weakly peaked near 17 meV and 9 meV.  $\Delta$$\chi''$($Q, \hbar\omega$) is strongly depleted at larger $Q$ and low energies indicating an approximate conservation of $\Delta$$\chi''$($Q, \hbar\omega$) with temperature, as expected.

 \begin{figure}[h]
 \includegraphics[width=85mm]{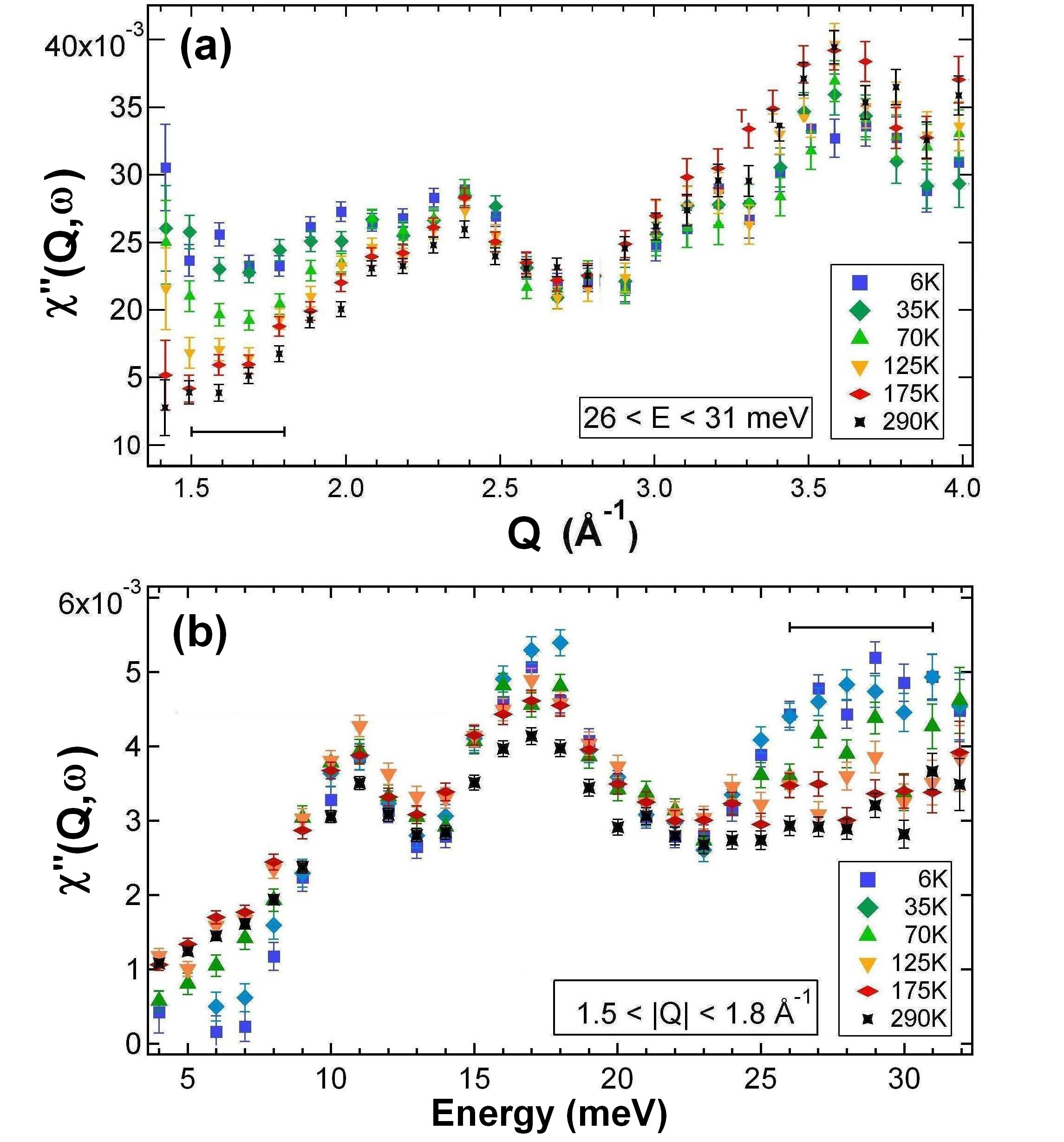}
 \caption{\label{} (a)  $\chi''$($Q, \hbar\omega$) plotted versus $Q$ for six temperatures, integrated in energy between 26 and 31 meV.  (b) $\chi''$($Q, \hbar\omega$) plotted versus energy  for six temperatures, integrated in $Q$ over the range 1.5 \text{\AA}$^{-1} <   Q < 1.8 $ \text{\AA}$^{-1}$.   The scattering centered on $\sim$28 meV exists only at low $Q<$ 2.5~$\AA^{-1}$ and at low $T<$ 125 K, and is therefore magnetic in origin and consistent with a weakly dispersive spin triplet excitation. }
\end{figure}

 \begin{figure}[h]
 \includegraphics[width=85mm]{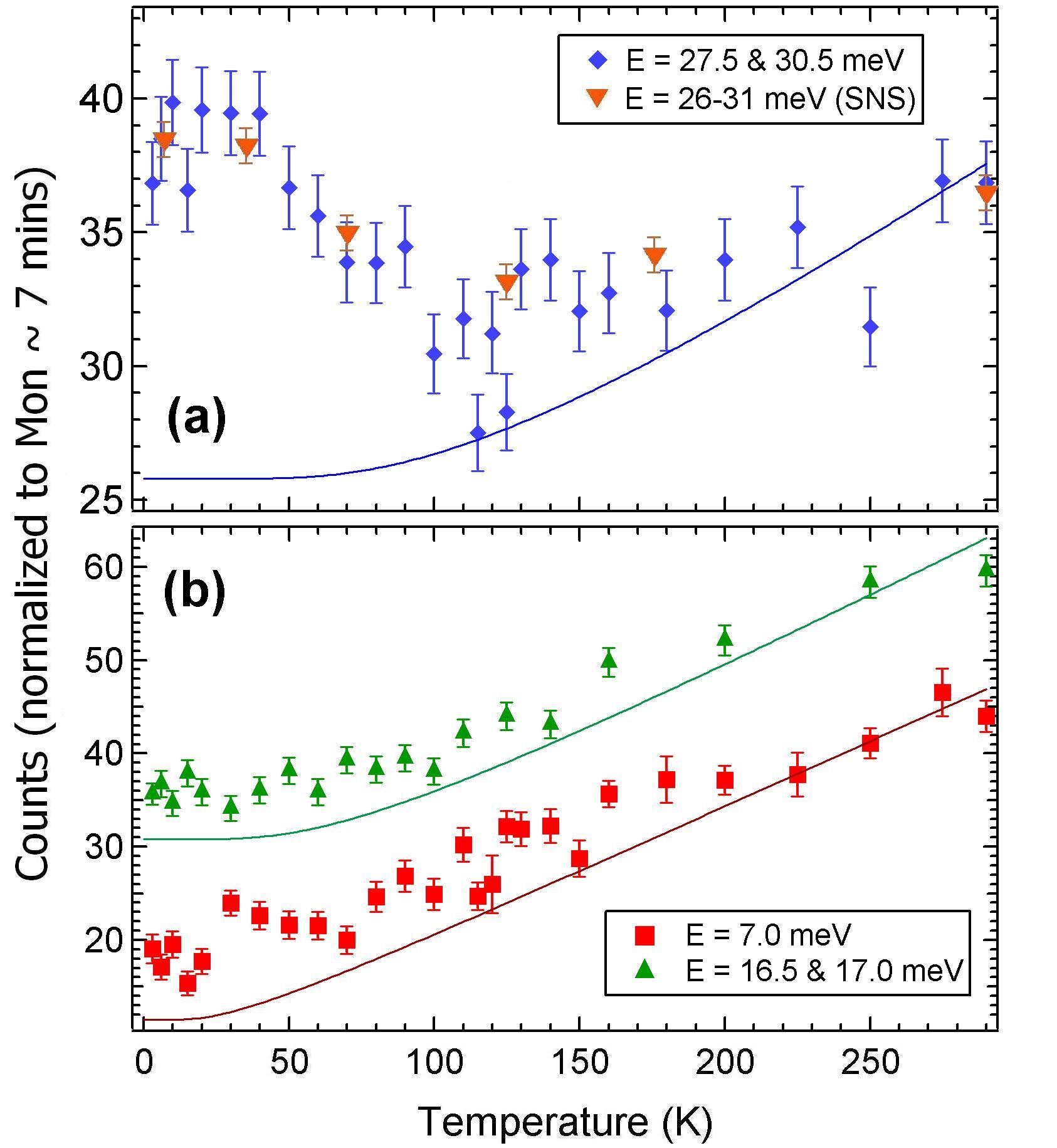}
 \caption{\label{}(a) Temperature dependence of the background-subtracted scattering intensity at $Q$=1.7 $\text{\AA}^{-1}$ at the average of 27.5 and 30.5 meV, collected with the C5 triple-axis spectrometer, showing a characteristic fall-off of the triplet intensity toward zero at $\sim$125~K; normalized SEQUOIA (SNS) data at 26-31 meV is included for reference.  (b) Temperature dependence of the background-subtracted intensity at 7 meV and 16.5$-$17 meV energy transfer.  The solid lines represent fits of the T$>$200~K data to the thermal occupancy factor.  Excess low-temperature scattering is attributed to either (a) the triplet excitation, or (b) magnetic states within the gap. }
 \end{figure}

Fig.~3 shows the $Q$ and $E$ dependence of $\chi''$($Q,\hbar\omega$), with emphasis on the magnetic signal at 28~meV.  Fig. 3(a) shows the $Q$-dependence of $\chi''$($Q,\hbar\omega$) integrated over 26 to 31~meV as a function of temperature, while Fig. 3(b) shows the energy dependence of $\chi''$($Q,\hbar\omega$) integrated in $Q$ from 1.5 to 1.8 $\text{\AA}^{-1}$.   As Fig. 3(b) shows $\chi''$($Q,\hbar\omega$), rather than the $\Delta$$\chi''$($Q,\hbar\omega$) shown in Fig. 2, peaks appear at energies corresponding to high optic and acoustic phonon density of states, near 17 and 11 meV.  Taken together, Figs. 3(a) \& 3(b) clearly show the 28 meV feature to be localized to low $Q$ $<$ 2.5 $\text{\AA}^{-1}$, and vanishing at temperatures above $\sim$125~K.  It is therefore magnetic in origin, and consistent with a weakly dispersive spin triplet excitation out of a singlet ground state. 

The detailed temperature dependence of the low$-Q$ scattering is shown in Fig. 4.  This data, taken with the C5 triple axis spectrometer at CNBC, Chalk River, shows background-subtracted scattering intensity at $Q$ = 1.7 \text{\AA}$^{-1}$ at three energy transfer values.   In Fig. 4(a) we show the detailed temperature dependence of the magnetic scattering at 27.5 and 30.5 meV (summed together).  Consistent with the SEQUOIA data, shown for comparison, this scattering falls off with increasing temperature, evolving to a slowly increasing, phonon-like background above $\sim$125 K.  We have modeled the high temperature ($T>$200~K) data with the Bose thermal occupancy factor, $[n(\omega)+1]$, which is plotted as solid lines on top of all three data sets in both figure panels.    Fig. 4(b) shows the temperature dependence of scattering at 7 meV and 16.75 meV; although scattering in this region is significantly affected by phonons, particularly the bands near 11 and 17 meV evident in Fig. 3(b), the temperature dependence of scattering at 7 meV and 16.75 meV do not fit to a purely Bose distribution, with a low-temperature excess consistent with magnetic scattering in the in-gap regime.  Taken together with the SEQUOIA data in Figs. 2 and 3, we have a clear and robust signature for two distributions of magnetic scattering at small $Q$: triplet excitations out of an exotic singlet ground state in Ba$_2$YMoO$_6$ with an energy gap of 28 meV, and paramagnetic-like scattering within this gap.  A paramagnetic state is recovered for $T > $125 K, with no obvious signs of a phase transition. 

While such spin excitations are unconventional, it is instructive to compare them to other $s$=$^1$$\!$/$_2$ singlet systems, particularly SrCu$_2$(BO$_3$)$_2$.   In this system, orthogonal singlet dimers decorate a square lattice, a 2D analogue to the pairs of orthogonal dimers which the $s$=$^1$$\!$/$_2$ moments on individual tetrahedra may constitute within Ba$_2$YMoO$_6$.  The triplet excitation in  SrCu$_2$(BO$_3$)$_2$ has a gap of $\sim$2.9 meV with little dispersion, and evolves without a phase transition at T $\sim$ 10~K; these dependences are analogous to those we report here for the 28 meV spin excitation in Ba$_2$YMoO$_6$, scaled down by roughly a factor of 10.  Furthermore, introduction  of small amounts of non-magnetic impurities in SrCu$_2$(BO$_3$)$_2$, with non-magnetic Mg$^{2+}$ substituting at the 2.5 $\%$ level for  $s$=$^1$$\!$/$_2$ Cu$^{2+}$ ions, gives rise to states within the singlet gap, broad in energy but weakly peaked at $\sim$ 66 and 35 $\%$ of the triplet gap  energy.  This is analogous to the continuum of magnetic scattering observed below $\sim$ 28 meV in Ba$_2$YMoO$_6$.  It is tempting to associate the magnetic states within the gap with the weak $\sim$ 3 $\%$ B-site disorder in our sample of Ba$_2$YMoO$_6$, although such in-gap magnetic states could also be intrinsic to the  $s$=$^1$$\!$/$_2$ FCC AF system in the presence of spin-orbit coupling.

While the magnetic excitation spectrum we observe in Ba$_2$YMoO$_6$ is exotic, containing two distributions of scattering, it is consistent with NMR measurements which indicated a low-temperature coexistence of paramagnetism and a collective singlet.  We also note that though the $\sim$28~meV energy gap we observe directly here is about a factor of two larger than that inferred from the NMR 1/$T_1$ temperature dependence of Aharen \textit{et al.} \cite{tomoko_1_2}, the temperature scale at which a conventional paramagnetic state is recovered is $\sim$125 K, within 15$\%$ of that seen in the NMR measurements; this is the likely origin of the 1/T$_1$ temperature dependence. 

To conclude, neutron spectroscopy directly reveals the spin-triplet excitation out of an exotic $s$=$^1$$\!$/$_2$ FCC spin-liquid singlet ground state in Ba$_2$YMoO$_6$.  Two INS experiments demonstrate the existence and temperature dependence of a gapped spin excitation at $\sim$28 meV with a bandwidth of $\sim$4 meV.  We also find a weakly-peaked continuum of magnetic states at lower energies within this gap, consistent with spin-polaron states associated with weak disorder.  No evidence of a phase transition to a disordered state is seen; rather, the magnetic signal weakens in intensity and disappears on a temperature scale of $\sim$125 K.  We hope that this work guides and informs a more comprehensive understanding of this little-studied class of frustrated quantum magnets. 

\begin{acknowledgments}
Work at McMaster University and Chalk River was supported by NSERC.  Research at Oak Ridge National Laboratory's Spallation Neutron Source was sponsored by the Scientific User Facilities Division, Office of Basic Energy Sciences, U. S. Department of Energy.  We acknowledge L.~Balents for illuminating discussions.
\end{acknowledgments}


%

\end{document}